\documentclass[journal]{IEEEtran}

\usepackage{cite}
\usepackage{amsmath,amssymb,amsfonts}
\usepackage{algorithmic}
\usepackage{textcomp}
\usepackage{mathtools}

\usepackage{tabularx, booktabs}
\usepackage{multirow, makecell}

\usepackage{graphicx}
\graphicspath{{./Figures/}}

\usepackage{epstopdf}
\usepackage{subfigure}
\usepackage{array}
\usepackage[ruled,vlined]{algorithm2e}
\usepackage{enumerate}
\usepackage{color}

\definecolor{Blue1}{rgb}{0.3,0,1}

\def\BibTeX{{\rm B\kern-.05em{\sc i\kern-.025em b}\kern-.08em
		T\kern-.1667em\lower.7ex\hbox{E}\kern-.125emX}}

\begin{document}
%
\markboth{IEEE Transactions on Smart Grid}%
{}

\title{Distributionally Robust Safety Filter for Learning-Based Control in Active Distribution Systems}
\author{Hoang~Tien~Nguyen,~\IEEEmembership{Graduate Student Member,~IEEE,}
	and~Dae-Hyun~Choi,~\IEEEmembership{Member,~IEEE}
\thanks{This work was supported in part by Basic Science Research Program through the National Research Foundation of Korea (NRF) funded by the Ministry of Education under Grant 2022R1F1A1062888 and in part by the NRF funded by the Korea government (MSIT) under Grant 2021R1A4A1031019.
}
\thanks{H.T. Nguyen and D.-H. Choi are with the School of Electrical and Electronics Engineering, Chung-Ang University, Dongjak-gu, Seoul 156-756, Korea (email: nguyentienhoangbk@gmail.com; dhchoi@cau.ac.kr). D.-H. Choi is the corresponding author of this paper.}
\thanks{© 20xx IEEE. Personal use of this material is permitted. Permission from IEEE must be
	obtained for all other uses, in any current or future media, including
	reprinting/republishing this material for advertising or promotional purposes, creating new
	collective works, for resale or redistribution to servers or lists, or reuse of any copyrighted component of this work in other works.}
}

\maketitle

\begin{abstract}
Operational constraint violations may occur when deep reinforcement learning (DRL) agents interact with real-world active distribution systems to learn their optimal policies during training.
This letter presents a universal distributionally robust safety filter (DRSF) using which any DRL agent can reduce the constraint violations of distribution systems significantly during training while maintaining near-optimal solutions.
The DRSF is formulated as a distributionally robust optimization problem with chance constraints of operational limits.
This problem aims to compute near-optimal actions that are minimally modified from the optimal actions of DRL-based Volt/VAr control by leveraging the distribution system model, thereby providing constraint satisfaction guarantee with a probability level under the model uncertainty.
The performance of the proposed DRSF is verified using the IEEE 33-bus and 123-bus systems.

\end{abstract}

\begin{IEEEkeywords}
Distributionally robust optimization, deep reinforcement learning, safe learning, safety filter, Volt/VAr control.
\end{IEEEkeywords}

%
\IEEEpeerreviewmaketitle

\section{Introduction}

\IEEEPARstart{D}{eep} reinforcement learning (DRL) has recently become a core technology for achieving reliable and efficient active distribution system operation under model uncertainties owing to its model-free and fast execution.
For example, DRL-based Volt/VAr control (VVC) achieves acceptable nodal voltage profiles by coordinating the voltage regulators and inverters of distributed energy resources, e.g., photovoltaic (PV) systems, without accurate knowledge of the distribution system model~\cite{zhang2020deep}.

However, the training of the DRL-based VVC agent via interactions with real-world distribution systems may be unsafe, leading to numerous operational constraint violations.
To address this safety problem, the optimal operation of the distribution system is modeled as a constrained Markov decision process problem~\cite{wang2019safe}, which is solved with a constrained soft actor--critic (CSAC) method to obtain the optimal policy under operational constraint satisfaction.
However, since the CSAC method ensures constraint satisfaction only after successful training of the agent, many constraint violations may occur during training.
A safety module was developed in our previous work~\cite{nguyen2022three} to modify the unsafe actions of the DRL agents and remove voltage violations completely during training; however, this module requires significant communication resources between the agent and distribution system to collect sensor measurements.

Many previous studies, including the aforementioned works, present model-free DRL-based VVC algorithms by assuming no knowledge of the distribution system model. However, distribution system is not a black box in the real world, and the distribution system model may be known, albeit not absolutely accurate or time-varying.
Evidently, knowledge of the distribution system models can help DRL methods achieve safe distribution system operations during training and deployment~\cite{chen2022reinforcement}.
Unfortunately, the uncertainty of the distribution system model may adversely impact DRL performance.
Recently, model uncertainties have been efficiently addressed using an advanced optimization method named distributionally robust optimization (DRO).
DRO aims to find an optimal and robust solution against uncertainty by considering all possible distributions of uncertain data within an ambiguity set~\cite{xie2021distributionally}.
Recently, the DRO method based on the Wasserstein ambiguity set is becoming a promising approach with the advantages of i) tractable reformulation, ii) finite sample guarantee, and iii) asymptotic consistency~\cite{mohajerin2018data}.
The Wasserstein-based DRO has been widely used in smart grid applications such as look-ahead economic dispatch~\cite{Poolla2020} and model predictive control for smart electric vehicle charging station~\cite{nguyen2023distributionally}.

To address both safety and model uncertainty in DRL-based distribution system operation, a universal distributionally robust safety filter (DRSF) is proposed herein that guarantees safe operation of the distribution system while interacting with any DRL-based controller.
In this letter, the proposed DRSF uses the DRO method to satisfy DRL-based VVC-induced operational constraints (e.g., voltage magnitude, line current, and power flow at the substation) with a probability level under the uncertainty of the distribution system model.
In addition, an approximate solution approach is introduced to improve the DRSF computation time.
Case studies are then performed using the IEEE 33-bus and 123-bus systems to verify the effectiveness of the proposed DRSF.

\section{Distributionally Robust Safety Filter}\label{sec:solution}
\subsection{Distribution System Model}
We consider a radial distribution system with a bus set $\mathcal{B}$ and line set $\mathcal{L}$ having the following DistFlow model~\eqref{eq:P balance}--\eqref{eq:line power}~\cite{farivar2013branch}:
\begin{align}
	&\sum_{i:(i,j)\in \mathcal{L}} \left(P_{ij} - r_{ij} l_{ij} \right) + P_j = \sum_{k:(j,k)\in \mathcal{L}} P_{jk},  \label{eq:P balance}\\
	&\sum_{i:(i,j)\in \mathcal{L}} \left(Q_{ij} - x_{ij} l_{ij} \right) + Q_j = \sum_{k:(j,k)\in \mathcal{L}} Q_{jk},  \label{eq:Q balance}\\
	&P_j = P^\text{pv}_j - P^\text{load}_j, \quad Q_j = Q^\text{pv}_j - Q^\text{load}_j \label{eq:bus power}\\
	&v_j = v_i - 2(r_{ij}P_{ij} + x_{ij}Q_{ij}) + (r_{ij}^2 + x_{ij}^2)l_{ij}, \label{eq:bus voltage}\\
	&v_i l_{ij} = P_{ij}^2 + Q_{ij}^2 \label{eq:line power}.
\end{align}
where $P_{ij}$($Q_{ij}$) is the real (reactive) power flow from bus $i$ to $j$; $r_{ij}$ ($x_{ij}$) is the resistance (reactance) of the line connecting buses $i$ and $j$; $l_{ij}$ is the squared current flowing from bus $i$ to $j$; $v_i$ is the squared voltage magnitude at bus $i$; $P^\text{pv}_j$ ($Q^\text{pv}_j$) is the real (reactive) power generation of the PV system at bus~$j$; $P^\text{load}_j$ ($ Q^\text{load}_j$) is the real (reactive) power consumption of the load at bus $j$. 
In this paper, we assume that the topology of the distribution system is known and fixed, and the line parameters $x_{ij}$ and $r_{ij}$ are known with inaccurate values because these values are difficult to measure accurately in practice. 
In the next section, we design a safety filter based on distributionally robust optimization to guarantee safe learning of learning-based controllers while considering the uncertainty in the line parameters. 

\begin{figure}
	\centering
	\includegraphics[width=3.2in]{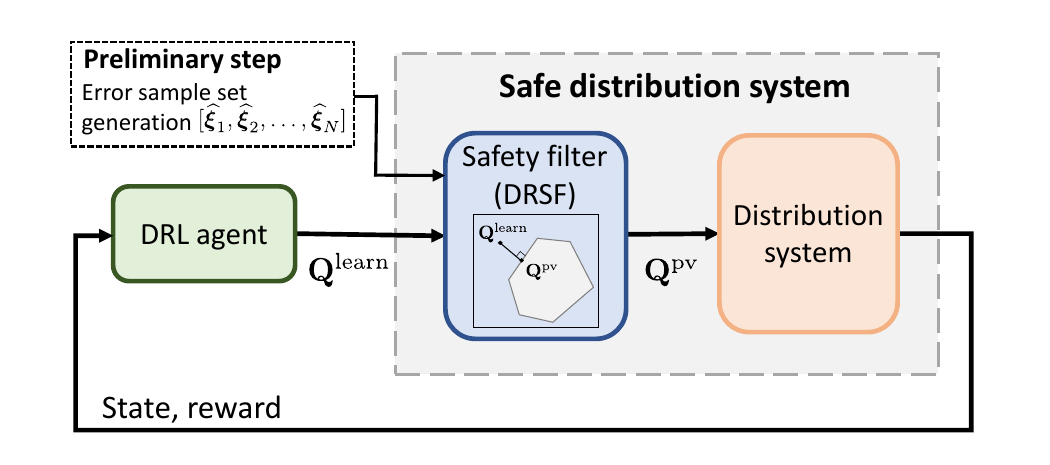}
	\vspace*{-0.15in}
	\caption{Structure of closed-loop distribution system controlled by a DRL agent with the proposed DRSF.}
	\label{fig:structure}
	\vspace*{-0.2in}
\end{figure}

\subsection{Mathematical Formulation of DRSF}
Fig.~\ref{fig:structure} depicts the structure of the closed-loop distribution system interacting with the DRL agent for VVC and the proposed DRSF.
Here, the primary goal of the DRSF is to modify the actions $\mathbf{Q}^\text{learn}$ (optimal yet unsafe reactive power generations of PV systems) of the DRL agent to ensure safe distribution system operation under model uncertainty while maintaining near-optimal actions $\mathbf{Q}^\text{pv}$.
Note that with the proposed DRSF, the distribution system can safely interact with \emph{any learning-based controller}.

The DRSF is formulated as a DRO problem, which aims to compute minimally modified  actions $\mathbf{Q}^\text{PV}$ from unsafe actions $\mathbf{Q}^\text{learn}$ by minimizing  the difference between them and the weighted penalty for the guarantee of the second-order cone program (SOCP) model relaxation exactness~\cite{farivar2013branch} in~\eqref{eq:obj} while satisfying the system operational constraints~\eqref{eq:four constraint}--\eqref{eq:cc p}:

\begin{align}
	& \min_{\mathbf{Q}^\text{pv}} J = \| \mathbf{Q}^\text{learn} - \mathbf{Q}^\text{pv} \| + \omega \sum_{(i,j) \in \mathcal{L}} l_{ij}\label{eq:obj} \\
	\text{s.t. }
	&\text{Eqn. } \eqref{eq:P balance}-\eqref{eq:bus voltage},\label{eq:four constraint} \\
	& \left\| \left[2 P_{ij},\, 2 Q_{ij},\, v_i - l_{ij} \right]^\top \right\| \leq (v_i + l_{ij})^2, \label{eq:cone constraint} \\
	& (P_i^\text{pv})^2 + (Q_i^\text{pv})^2 \leq (S_i^\text{pv})^2, \label{eq:PV limit}\\	
	&\hspace*{-0.1in} \inf_{\mathbb{P} \in \mathcal{P}_N(\epsilon)} \mathbb{P} \left[\underline{V}^2 \leq v_i + \Delta v_i \leq \overline{V}^2, \, i \in \mathcal{N} \right] \geq 1- \alpha, \label{eq:cc v}\\
	&\hspace*{-0.1in} \inf_{\mathbb{P} \in \mathcal{P}_N(\epsilon)} \mathbb{P} \left[l_{ij} + \Delta l_{ij} \leq \overline{I}^2, \, (i,j) \in \mathcal{L} \right] \geq 1- \alpha, \label{eq:cc l}\\
	&\hspace*{-0.1in} \inf_{\mathbb{P} \in \mathcal{P}_N(\epsilon)} \mathbb{P} \left[ P_0^2 + Q_0^2 + \Delta S_0 \leq \overline{S}_0^2 \right]  \geq 1- \alpha. \label{eq:cc p}
\end{align}
The quadratic constraints~\eqref{eq:line power} are relaxed to second-order cone constraints~\eqref{eq:cone constraint}.
Constraints~\eqref{eq:PV limit} denote the reactive power limit of the PV system at bus $i$ with real power $P_i^\text{pv}$ and apparent power $S_i^\text{pv}$.
Chance constraints~\eqref{eq:cc v}--\eqref{eq:cc p} represent the system operational constraints where voltage magnitude error $\Delta v_i$, line current error $\Delta l_{ij}$, and apparent power flow error $\Delta S_0$ at the substation stem from the distribution system model uncertainties.\footnote[1]{The time subscripts of the variables are omitted for notational brevity}
These chance constraints must be satisfied with a probability of at least $1-\alpha$ under all probability distributions of the uncertainties within the Wasserstein ambiguity set $\mathcal{P}_N(\epsilon)$.
The Wasserstein ambiguity set for a random vector $\boldsymbol{\xi}$ is constructed as a Wasserstein ball of radius $\epsilon$ centered at an empirical distribution $\widehat{\mathbb{P}}_N$ as follows:
\begin{equation}\label{eq:ambiguity_set}
	\mathcal{P}_N(\epsilon) = \left\{ \mathbb{P} \in \mathcal{M}(\Xi) : d_W(\mathbb{P}, \widehat{\mathbb{P}}_N) \leq \epsilon \right\}
\end{equation}
where the empirical distribution $\widehat{\mathbb{P}}_N$ is expressed as $\widehat{\mathbb{P}}_N := \frac{1}{N} \sum_{s=1}^N \delta_{\widehat{\boldsymbol{\xi}}_s}$ with $N$ historical samples $\widehat{\boldsymbol{\xi}}_s$ (where $\delta_{\widehat{\boldsymbol{\xi}}_s}$ is the unit point mass at $\widehat{\boldsymbol{\xi}}_s$); $\mathcal{M}(\Xi)$ is the probability space of all probability distributions $\mathbb{P}$ supported on the uncertainty set $\Xi$; $d_W$ is the Wasserstein metric (see~\cite{mohajerin2018data}, \cite{nguyen2023distributionally}).

\subsection{Approximate Solution by Robust Optimization}

For notational simplicity, the $D\times 1$ random vector $\boldsymbol{\xi} \in \mathbb{R}^D$ represents the voltage error vector $\mathbf{\Delta v} = [ \Delta v_i \vert i \in \mathcal{B}]^\top$, current error vector $\mathbf{\Delta l} = [ \Delta l_{ij} \vert (i,j) \in \mathcal{L}]^\top$, and apparent power flow error $\Delta S_0$, respectively.
As shown in Fig.~\ref{fig:structure}, an error sample set $[\boldsymbol{\widehat{\xi}}_1, \boldsymbol{\widehat{\xi}}_2,\dots, \boldsymbol{\widehat{\xi}}_N]$ of the random error vectors is generated as a preliminary step prior to the execution of the DRSF.
In general, the error sample set can be obtained by subtracting the power flow solution of the uncertain distribution system model from the corresponding actual measurements of voltage, current, and substation real power flow.
The error sample set is used to compute the empirical distribution $\widehat{\mathbb{P}}_N$ of the random error vectors in the Wasserstein ambiguity set~\eqref{eq:ambiguity_set}, thereby characterizing their uncertainties in the DRSF problem.

To solve the DRSF efficiently, we adopt an approximate solution approach that includes calculation of the uncertainty bounds and robust optimization counterpart~\cite{margellos2014road}. The procedure comprises the following two steps:

\textit{Step 1: Distributionally Robust Bound Determination} \\
The distributionally robust bounds $[\overline{\boldsymbol{\xi}}, \underline{\boldsymbol{\xi}}]$ of the random error vector $\boldsymbol{\xi}$ are determined by solving the following optimization problem:
\begin{align}
	& \min_{\overline{\boldsymbol{\xi}}, \underline{\boldsymbol{\xi}} \in \mathbb{R}^D} \quad \vert \overline{\boldsymbol{\xi}} - \underline{\boldsymbol{\xi}} \vert \\
	\text{s.t. }
	& \underline{\boldsymbol{\xi}} \leq \mathbf{0} \leq \overline{\boldsymbol{\xi}} \\
	& \inf_{\mathbb{P} \in \mathcal{P}_N(\epsilon)} \mathbb{P} \left[ \boldsymbol{\xi} \in \left[ \underline{\boldsymbol{\xi}},\overline{\boldsymbol{\xi}} \right] \right] \geq 1 -\alpha.
\end{align}
This problem can be reformulated to as a tractable mixed-integer programming problem as follows \cite{xie2021distributionally}:
\begin{align}
	& \min_{\overline{\boldsymbol{\xi}}, \underline{\boldsymbol{\xi}} \in \mathbb{R}^d} \quad \vert \overline{\boldsymbol{\xi}} - \underline{\boldsymbol{\xi}} \vert \\
	\text{s.t. }
	& \underline{\xi}_k \leq 0 \leq \overline{\xi}_k, \label{eq:bound conditions}\\
	& \alpha \gamma - \epsilon v   \geq \frac{1}{N} \sum_{s=1}^N z_s ,\\
	& \gamma -z_s \leq r_s , \\
	& r_j \leq \overline{\xi}_k - \widehat{\xi}_{k,s} + M_s (1 - y_s), \\
	& r_j \leq - \underline{\xi}_k + \widehat{\xi}_{k,s} + M_s (1 - y_s), \\
	& r_j \leq M_s y_s ,  \\
	& v \geq 1, \gamma \geq 0, r_s \geq 0, z_s \geq 0, y_s \in \{0, 1\}.
\end{align}
where $k$ and $s$ are the indices of the elements and samples of the random error vector, respectively; $M_s$ is the big-M coefficient for relaxation; $v, \gamma, r_s, z_s, y_s$ are auxiliary variables.

\textit{Step 2: Robust Optimization Counterpart}

The distributionally robust bounds calculated from Step 1 are used to reformulate the chance constraints~\eqref{eq:cc v}--\eqref{eq:cc p}, which then become deterministic constraints as follows:
\begin{align}
	& \underline{V}^2 \leq v_i + \underline{\Delta v_i}, ~~v_i + \overline{\Delta v_i} \leq \overline{V}^2, \label{eq:robust cons 1}\\
	& l_{ij} + \overline{\Delta l_{ij}} \leq \overline{I}^2, \label{eq:robust cons 2}\\
	& P_0^2 + Q_0^2 + \overline{\Delta S_0} \leq \overline{S}_0^2. \label{eq:robust cons 3}
\end{align}

Note that the computational complexities of \eqref{eq:robust cons 1}--\eqref{eq:robust cons 3} are small and unchanged even with increasing numbers of error samples $N$. This is  because only the upper and lower error bounds information is used to solve the DRSF problem.

\section{Simulation results}\label{sec:case study}

The performance of the proposed DRSF was verified using the IEEE-33 bus and 123-bus systems~\cite{nguyen2022three}. The parameters and locations of the PV systems were obtained from~\cite{nguyen2022three}.
The limits of the voltage magnitude, line current, and power flow at the substation were chosen as $[\underline{V}, \overline{V}]=[0.95,1.05]$ p.u., $\overline{I}=3.46$ p.u., and $\overline{S}_0 = 3.46$ p.u., respectively.
The parameters of the DRSF problem were set as follows: $N=50$, $\alpha=0.1$, $\epsilon=0.01$, and  $\omega=10^{-5}$. For model uncertainty, the distribution system line parametedrs are assumed to be time-varying, where the deviations of the actual parameters from the nominal ones follow a normal distribution $\mathcal{N}(0,0.3)$.
To generate the error sample set for the DRSF execution, the power flow solution is first obtained from the inaccurate distribution system model with the aforementioned uncertain line parameters and random actions of the DRL agent. Then, the inaccurate model-based power flow solution is subtracted from the power flow solution of the accurate distribution system model, which generates the error samples.

\begin{figure}[t!]
	\centering
	\includegraphics[width=3.4in]{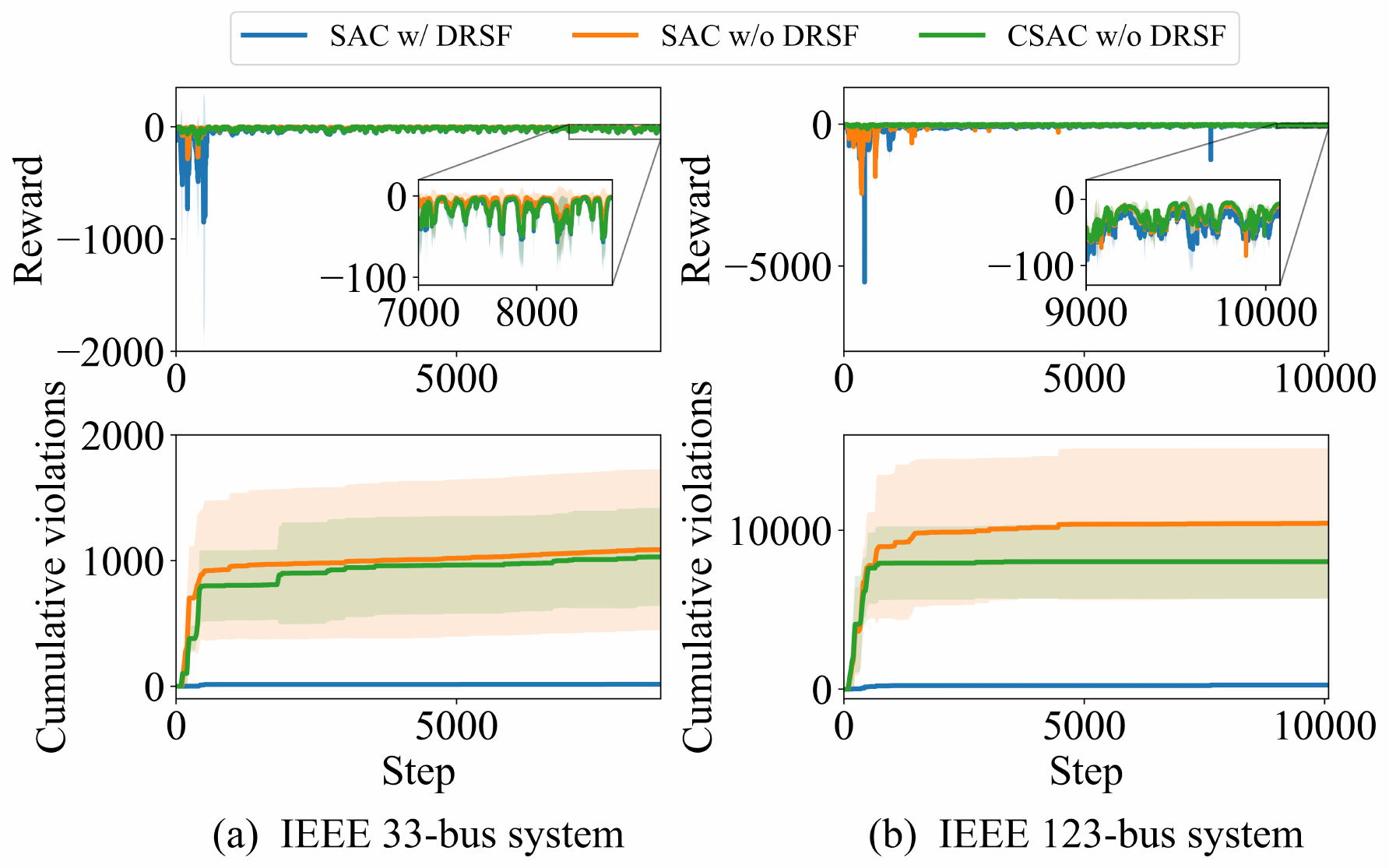}
	\caption{Reward convergence and cumulative constraint violations of the DRL methods during training.}
	\label{fig:training result}
\end{figure}

\subsection{Safety Performance of the Proposed DRSF}

The safety performance of the soft actor-critic (SAC)-based VVC with the proposed DRSF is compared with two DRSF-free DRL methods: i) SAC-based VVC and ii) CSAC-based VVC~\cite{wang2019safe}.
The reward function of SAC-based VVC with the DRSF is defined as
\begin{equation}\label{eq:reward_function}
	r = -\omega_1 \| \mathbf{Q}^\text{learn} - \mathbf{Q}^\text{pv} \| - \omega_2 P^\text{loss}
\end{equation}
where $\| \mathbf{Q}^\text{learn} - \mathbf{Q}^\text{pv} \|$ is the action deviation obtained from the DRSF problem; $P^\text{loss}$ is the real power loss; and $\omega_1$ and $\omega_2$ are positive weights with $(\omega_1, \omega_2) =  (2000, 1000) \text{ and } (3000, 1000)$ for the IEEE 37-bus and 123-bus, respectively.
The state and action of all DRL methods are defined as the same ones in~\cite{nguyen2022three}.
Fig.~\ref{fig:training result} shows the reward convergence and cumulative constraint violations for three DRL methods during training in the IEEE 33-bus and 123-bus systems.
Note from this figure that the training curves of all three methods converge to almost the same reward.

\begin{figure}
	\centering
	\includegraphics[width=3.5in]{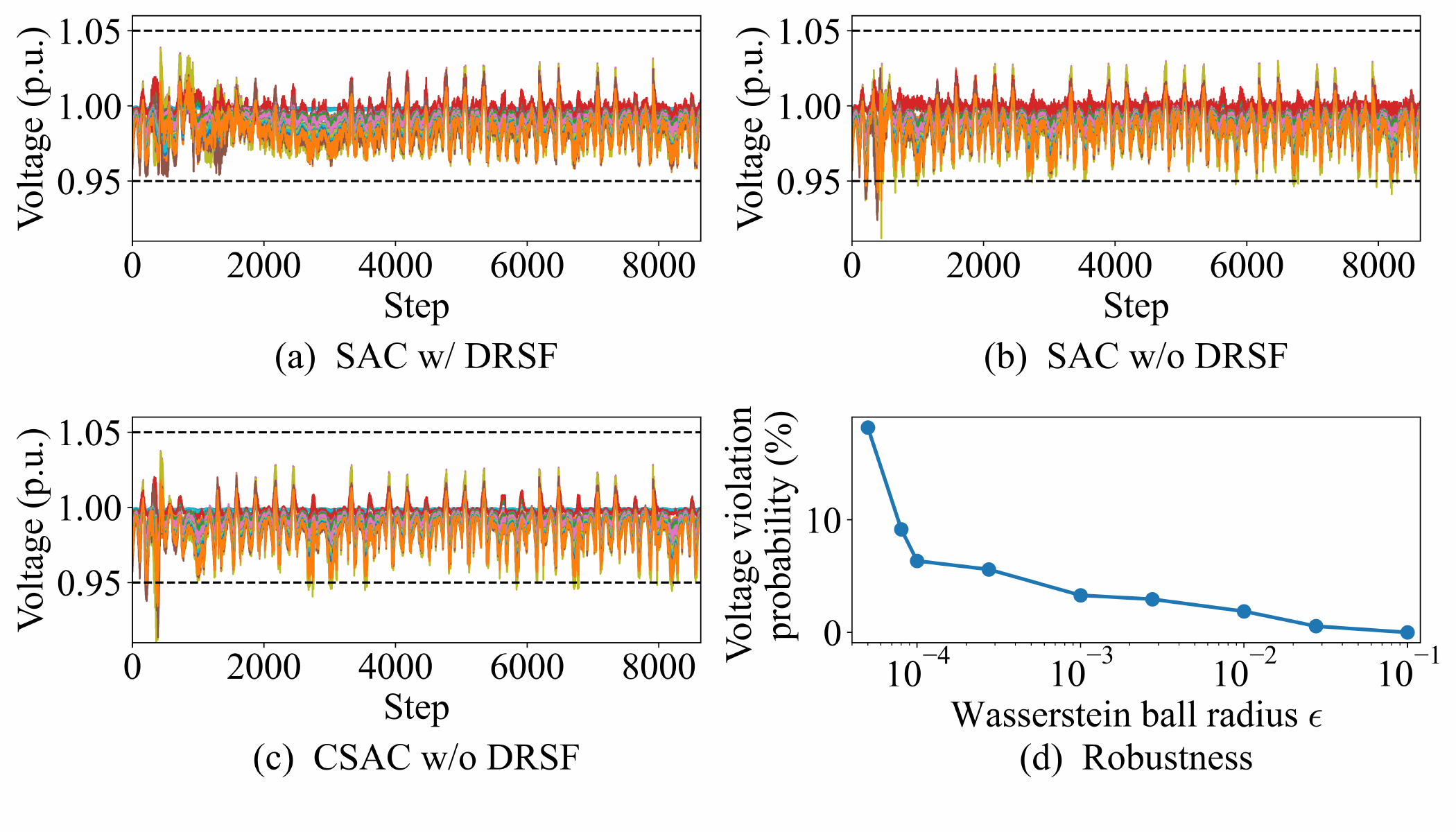}
	\caption{Voltage profiles of the DRL methods during training.}
	\label{fig:training voltage and robustness}
\end{figure}

However, note from Fig.~\ref{fig:training result}(a) that the SAC with DRSF yields a small number of constraint violations with a mean of 5.3 during training in the IEEE 33-bus system, whereas the SAC and CSAC without DRSF show large numbers of constraint violations with mean values of 1086.3 and 1029, respectively.
Similar observations are noted for the IEEE 123-bus system, as shown in Fig.~\ref{fig:training result}(b).
Figs.~\ref{fig:training voltage and robustness}(a)--(c) show the voltage magnitudes of all buses for these three DRL methods during training in the IEEE 33-bus system. Note from these figures that many voltage violations occur for the SAC and CSAC without DRSF, while no voltage violations occur for the SAC with DRSF.

Table~\ref{tab:test result} reports the test results of the trained policies of the three DRL methods using data from ten days in terms of the numbers of constraint violations and average real power losses.
After being trained successfully, all three methods remove the constraint violations completely and generate similar average real power losses.
This observation demonstrates that the action modification via the DRSF has negligible impact on the DRL performance for reduction of the real power loss.
\begin{table}
	\centering
	\caption{Test results of the DRL methods on 33-bus and 123-bus systems}
	\label{tab:test result}
		\begin{tabular}{>{\centering}p{2.1cm}  >{\centering}p{0.9cm} >{\centering}p{1.4cm} >{\centering}p{0.9cm} >{\centering\arraybackslash}p{1.4cm} }
			\toprule
			\multirow{3}{*}{Method}& \multicolumn{2}{c}{33-bus} & \multicolumn{2}{c}{123-bus} \\
			\cmidrule{2-5}
			& Constraint & Avg. power & Constraint & Avg. power    \\
			& violations & loss (kW) &  violations & loss (kW)  \\
			\midrule
			\textbf{SAC w/ DRSF } & \textbf{0}   &  \textbf{26.150} & \textbf{0} &  \textbf{22.737}	\\
			SAC w/o DRSF & 0 & 25.974 & 0 & 21.942\\
			CSAC w/o DRSF & 0 & 26.192 &0&22.819 \\
			No VVC & 6886 & 43.656 &27441 & 33.201 \\
			\bottomrule
		\end{tabular}
	\end{table}

\subsection{Robustness and Computation Time}

Fig.~\ref{fig:training voltage and robustness}(d) shows the impact of Wasserstein ball radius $\epsilon$ for the DRSF on the voltage violation probability. Note from this figure that a larger $\epsilon$ yields a smaller voltage violation probability, which ensures a robust solution against the model uncertainty.

The average solving times of the DRSF problem in the IEEE 33-bus and 123-bus systems were 0.01534 s and 0.04625 s, respectively, which are small compared to the VVC scheduling time (order of minutes).
Therefore, the proposed DRSF can be integrated with real-time VVC for safe operation of the distribution system.

\section{Conclusions}\label{sec:conclusions}

This letter proposes a universal DRO-based DRSF that enables any DRL controller to ensure safe distribution system operation during training while handling the uncertainties of the distribution system model associated with voltage magnitude, line current, and real power flow at the substation.
An approximate solution approach based on determination of the distributionally robust bounds of the uncertainties is adopted to solve the DRSF problem efficiently.
The simulation results with the IEEE 33-bus and 123-bus systems confirm the effectiveness of the proposed DRSF in terms of DRL-based VVC-induced constraint violations and real power losses as well as computation times.

\bibliographystyle{IEEEtran}
\bibliography{references}

\begin{thebibliography}{10}
\providecommand{\url}[1]{#1}
\csname url@samestyle\endcsname
\providecommand{\newblock}{\relax}
\providecommand{\bibinfo}[2]{#2}
\providecommand{\BIBentrySTDinterwordspacing}{\spaceskip=0pt\relax}
\providecommand{\BIBentryALTinterwordstretchfactor}{4}
\providecommand{\BIBentryALTinterwordspacing}{\spaceskip=\fontdimen2\font plus
\BIBentryALTinterwordstretchfactor\fontdimen3\font minus
  \fontdimen4\font\relax}
\providecommand{\BIBforeignlanguage}[2]{{%
\expandafter\ifx\csname l@#1\endcsname\relax
\typeout{** WARNING: IEEEtran.bst: No hyphenation pattern has been}%
\typeout{** loaded for the language `#1'. Using the pattern for}%
\typeout{** the default language instead.}%
\else
\language=\csname l@#1\endcsname
\fi
#2}}
\providecommand{\BIBdecl}{\relax}
\BIBdecl

\bibitem{zhang2020deep}
Y.~Zhang, X.~Wang, J.~Wang, and Y.~Zhang, ``Deep reinforcement learning based
  volt-var optimization in smart distribution systems,'' \emph{IEEE
  Transactions on Smart Grid}, vol.~12, no.~1, pp. 361--371, Jan. 2020.

\bibitem{wang2019safe}
W.~Wang, N.~Yu, Y.~Gao, and J.~Shi, ``Safe off-policy deep reinforcement
  learning algorithm for volt-var control in power distribution systems,''
  \emph{IEEE Transactions on Smart Grid}, vol.~11, no.~4, pp. 3008--3018, Jun.
  2019.

\bibitem{nguyen2022three}
H.~T. Nguyen and D.-H. Choi, ``Three-stage inverter-based peak shaving and
  volt-var control in active distribution networks using online safe deep
  reinforcement learning,'' \emph{IEEE Transactions on Smart Grid}, vol.~13,
  no.~4, pp. 3266--3277, Jul. 2022.

\bibitem{chen2022reinforcement}
X.~Chen, G.~Qu, Y.~Tang, S.~Low, and N.~Li, ``Reinforcement learning for
  selective key applications in power systems: Recent advances and future
  challenges,'' \emph{IEEE Transactions on Smart Grid}, vol.~13, no.~4, pp.
  2935--2958, Jul. 2022.

\bibitem{xie2021distributionally}
W.~Xie, ``On distributionally robust chance constrained programs with
  wasserstein distance,'' \emph{Mathematical Programming}, vol. 186, no.~1, pp.
  115--155, Mar. 2021.

\bibitem{mohajerin2018data}
P.~Mohajerin~Esfahani and D.~Kuhn, ``Data-driven distributionally robust
  optimization using the wasserstein metric: Performance guarantees and
  tractable reformulations,'' \emph{Mathematical Programming}, vol. 171, no.~1,
  pp. 115--166, Sep. 2018.

\bibitem{Poolla2020}
B.~K. Poolla, A.~R. Hota, S.~Bolognani, D.~S. Callaway, and A.~Cherukuri,
  ``{Wasserstein distributionally robust look-ahead economic dispatch},''
  \emph{IEEE Transactions on Power Systems}, vol.~36, no.~3, pp. 2010--2022,
  May 2020.

\bibitem{nguyen2023distributionally}
H.~T. Nguyen and D.-H. Choi, ``{Distributionally Robust Model Predictive
  Control for Smart Electric Vehicle Charging Station with V2G/V2V
  Capability},'' \emph{IEEE Transactions on Smart Grid}, early access,
  doi:10.1109/TSG.2023.3263470, 2023.

\bibitem{farivar2013branch}
M.~Farivar and S.~H. Low, ``{Branch flow model: Relaxations and
  convexification—Part I},'' \emph{IEEE Transactions on Power Systems},
  vol.~28, no.~3, pp. 2554--2564, Aug. 2013.

\bibitem{margellos2014road}
K.~Margellos, P.~Goulart, and J.~Lygeros, ``On the road between robust
  optimization and the scenario approach for chance constrained optimization
  problems,'' \emph{IEEE Transactions on Automatic Control}, vol.~59, no.~8,
  pp. 2258--2263, Aug. 2014.

\end{thebibliography}

\end{document}